\documentstyle[psfig,aps,epsf,amsmath,amssymb]{revtex}

\newcommand{\Wet}{W\!e_{\theta}}

\newcommand{\boldtau}{\mbox{\boldmath $\tau$}}

\newcommand{\boldv}{\mbox{\boldmath $v$}}
\newcommand{\boldalpha}{\mbox{\boldmath$\alpha$}}
\title{Solitary coherent  structures in viscoelastic shear flow:
computation and mechanism}
\date{\today}
\author{K. Arun Kumar and Michael D. Graham}
\address{Department of Chemical Engineering and Rheology Research
Center,
University of Wisconsin-Madison,
Madison, WI 53706-1691.}
\begin{document}
\maketitle
\thispagestyle{empty}
\section*{Abstract}
Starting from stationary bifurcations in  Couette-Dean flow, we
compute nontrivial stationary solutions in inertialess viscoelastic
circular 
Couette flow. These solutions are strongly localized vortex pairs,
 exist at arbitrarily large wavelengths, and show
hysteresis in the Weissenberg 
number, similar to experimentally observed ``diwhirl'' patterns. Based
on the computed velocity and stress fields,  
we elucidate a  heuristic, fully nonlinear mechanism for these
flows. We
propose that  these localized, fully nonlinear
structures
comprise fundamental building blocks for complex spatiotemporal
dynamics in the flow of elastic liquids. 

\vspace*{0.1in}

\noindent{\bf PACS.} 47.54.+r Pattern formation - 47.50.+d Non-Newtonian fluid
flows - 83.50.Ax Steady shear flows - 83.85.Pt Flow computation.

\section*{} 
Flow instabilities and nonlinear dynamics 
have long been recognized to
occur in flows of viscoelastic polymer melts and
solutions~\cite{denn76,larson92,shaqfeh96rev}.  An important 
breakthrough, which has led to increasing recent attention to these
phenomena, was made by Larson, Shaqfeh and Muller~\cite{lsm90}, who
discovered that circular Couette flow of a viscoelastic liquid
undergoes an instability loosely analogous to the classical
Taylor-Couette instability of Newtonian liquids, but driven solely by
elasticity - the instability is present at zero Taylor (Reynolds)
number. They also showed
that a linear stability analysis of a simple fluid model predicts
instability and elucidated the basic elasticity-driven mechanism of
the instability. In particular, although many researchers had studied
the effects of viscoelasticity on the Newtonian (inertial)
Taylor-Couette instability, both
experimentally~\cite{elata66,roisman69,denn72,giesekus72,joseph74},
and 
theoretically~\cite{datta64,walters64,elata66,roisman69,denn72,lange00},
these workers were the first to demonstrate an
inertialess, purely elastic mechanism for instability in a viscometric
flow.
More recent observations have revealed a wealth of
interesting 
dynamics in this flow as well as other simple
flows~\cite{shaqfeh96rev,steinberg97,muller99,steinberg00}; one set of
experimental observations of particular 
interest was made by Groisman and Steinberg~\cite{steinberg97}, who found
long-wavelength, stationary axisymmetric vortex pair structures in
inertialess viscoelastic flow in the circular Couette geometry.  There
are two interesting aspects to these observations: (1) isothermal  linear
stability analysis in this geometry never predicts bifurcation of
stationary states (in contrast to the classical Taylor-Couette case)
and (2) the observations suggest that these vortex pairs, which
Groisman and Steinberg dubbed ``diwhirls'',  can exist in isolation --
there does not seem to be a selected axial wavelength for this pattern. 
These considerations motivate the present computational study, which 
addresses the following questions: (1) Do isolated branches of
stationary solutions exist in a simple model of a viscoelastic fluid
in the circular Couette geometry? (2) If so, what are the spatial
structures of these?  Are they localized? (3) A nonlinear
self-sustaining mechanism must be present for such patterns to exist.
Can the computations help us elucidate it?

We address these questions by fully nonlinear computations of the
branching behavior of an inertialess isothermal FENE dumbbell fluid in
the circular Couette geometry.  Introduction of an azimuthal body force as
an additional parameter allows the possibility of accessing steady
solution branches that would be isolated in the parameter space of
Couette flow.  We find in short that the answers to all of the above
questions is yes: isolated solution branches consisting of spatially
localized coherent structures have been computed for the first time in a
purely elastic flow, in very good agreement with the experimental
observations, and a heuristic, fully 
nonlinear mechanism has been  elucidated.

We consider the flow of an inertialess polymer solution in the annulus
between two concentric  cylinders. The inner
cylinder has radius $R_1$ and the 
outer cylinder has radius $R_2$. The flow is assumed to periodic in
the axial direction, and we denote the period, nondimensionalized with
the gap width $R_2-R_1$,  by $L$.  
The fluid has a relaxation time
$\lambda$; the polymer and solvent contributions to the viscosity are denoted
respectively by $\eta_p$ and $\eta_s$, with the ratio $\eta_s/\eta_p$
denoted by $S$. 
The flow is created by a combination of the motion of the inner
cylinder at a velocity $\Omega R_1$ and by the application of an
azimuthal pressure gradient $K_\theta=\partial P/\partial \theta$. The
equations governing the flow are the momentum and mass
conservation equations, and the FENE-P constitutive
equation~\cite{dpl2}, which models polymers as beads connected by
finitely extensible springs. These may be written in dimensionless
form as
\begin{equation}
\nabla \cdot {\boldtau} - \nabla {p} + W\!e_{\theta} {S}
\nabla^2 {\boldv} = 0,
\label{momentum}
\end{equation}
\begin{equation}
\nabla \cdot \boldv =0,
\label{continuity}
\end{equation}
\begin{equation}
\Wet\,\left(\frac{D\boldalpha}{D
t}-\left\{\boldalpha\cdot\nabla\boldv\right\}^\dagger-\left\{\boldalpha\cdot\nabla\boldv\right\}\right)
+\left( \frac{\boldalpha}{(1-{\rm
tr}(\boldalpha)/b)}-{\bf I}\right)=0,
\label{constitutive}
\end{equation}
where  $\boldv$ is the velocity, $p$ is the pressure, $\boldalpha$ is
the ensemble average of the conformation tensor, and 
$\boldtau=\boldalpha/(1-{\rm tr}(\boldalpha)/b)-{\bf I}$  is
the polymer stress tensor, with $\sqrt{b}$ being 
a dimensionless measure of the maximum extensibility of the
dumbbells. The Weissenberg number, $\Wet$  is the product of
the polymer 
relaxation time and a characteristic shear rate, which we take to be
the shear rate at the outer cylinder for an Oldroyd-B fluid ($1/b=0$) flowing
through the geometry. Other important parameters are the dimensionless
gap width, $\epsilon=(R_2-R_1)/R_2$, and  $\delta$, which
measures the relative importance  of the pressure gradient as the
driving force for the flow, given by
\begin{equation}
\delta=\frac{-K_{\theta}\epsilon^2
R_2/(2\,(\eta_p+\eta_s))}{(1-\epsilon)R_2\Omega - K_{\theta}\epsilon^2
R_2/(2\,(\eta_p+\eta_s))},
\label{delta_defn}
\end{equation}
so that $\delta=0$ is circular Couette
flow and $\delta=1$ is Dean flow (pressure driven flow in a curved
channel). The velocity satisfies no slip boundary conditions on the
walls of the cylinder. We compute
steady, axisymmetric solutions to the 
governing equations using a 
 spectral element scheme~\cite{patera89} with  Galerkin weighting
 on the conservation equations and 
 streamline-upwind/Petrov-Galerkin weighting~\cite{hughes82,crochet87} on the
 constitutive equations. We use a branch  switching technique
 to compute starting points on non-trivial branches, and
 pseudo-arclength continuation to trace out solution curves
in parameter space~\cite{seydel94}. The linear systems in the Newton
iteration are solved using GMRES
\cite{gmres}, with a modified dual threshold
preconditioner~\cite{saad96,graham00}.

We restrict our
attention to parameters which are close to those used in the
experiments by Groisman and
Steinberg~\cite{steinberg97}. Specifically, we choose 
parameter values $S=1.2$, and $\epsilon=0.2$.  The Reynolds numbers in
their experiments were $\stackrel{<}{\sim} 1$, so our neglect of
inertia corresponds well to their experiments. In  our computations,
we find that the most 
interesting behavior occurs only for large values of $b$, so for  the 
results we present below, we choose a value of $1830$.

In the Couette flow case ($\delta=0$), there is no bifurcation from the
viscometric base state to a stationary nontrivial state -- if they exist,
any branches of nontrivial stationary states in this case are isolated.  So
to pursue such solutions, an indirect approach must be taken.  We use the
fact that in pure pressure-driven flow ($\delta=1$), stationary bifurcations
do exist -- by continuity these bifurcations also exist for $\delta<1$ and
we search for a route to the Couette case by exploring the evolution of these
bifurcating branches.  Specifically, we traced out a stationary branch
originating at $\delta=0.576$, $L=2.71$, $\Wet=25.15$, then increased
$L$ to $3.08$ while holding the values of the other parameters fixed,
then reduced $\delta$ downward, finding that this
solution branch persisted at $\delta=0$.  We have thereby shown that there is
indeed an isolated branch of nontrivial stationary patterns in the
circular Couette flow
geometry. If we now decrease $L$, we find a
 turning point at $L=2.94$, so these solutions do not exist for
 wavelengths shorter than this. On the other hand, when we increase
$L$, we find that the solutions exist at arbitrarily large
wavelength. Figure 1 shows the contours of  the
 streamfunction and $\alpha_{\theta\theta}$ at $L=110.89$ ( i.e., the
axial wavelength is more than 100 times times the gap width!) The
core of the pattern (which has been magnified for clarity)  is a region
of very strong inflow, surrounded 
by regions of much weaker outflow. The $\alpha_{\theta\theta}$
field shows an even more intense localization at the core 
and requires the use of a very fine spectral element mesh to
resolve. The strong localization at the core and the  asymmetry
between inflow and outflow are features shared by the experimentally
observed diwhirls~\cite{steinberg97}, which indicates that we are
capturing the physics behind the structures they observe. 
Away from the core, the flow field 
is almost pure circular Couette flow. In obtaining solutions at very
large values of $L$, we exploited the lack of axial
variation away from the  center and the fact that the characteristic
length of the localization near the core does not change at large
wavelengths  by using the same mesh as at smaller
values of $L$ close to the 
center, and simply increasing the lengths of the spectral elements bordering
the axial edges.

 Figure 2 shows the results of
 continuation in $\Wet$ for 
 three different values of $L$. In each case, we see a turning
 point in $\Wet$. It  is located at $\Wet=24.97$ for
 $L=3.08$, at 
 $\Wet=23.37$ for $L=4.72$, and at $\Wet=23.55$ for
 $L=9.11$. The point at which the diwhirls lose existence is
 much lower than the critical $\Wet$ for linear instability at that
 wavelength, so that the overall bifurcation structure shows a
 hysteretic character, consistent with experiment.

A natural question regarding these new flow states is their stability.
We have performed linear stability analyses with respect to axisymmetric
perturbations, for selected parameter values.  These results will be
detailed elsewhere~\cite{graham00}, but the main result is simple and
interesting: there is 
an unstable complex conjugate pair of eigenvalues, corresponding to
instability with respect to oscillatory disturbances.  The destabilizing
disturbance, however, has significant amplitude only near the ends of 
the domain, where the flow is essentially Couette flow (which is 
linearly unstable at
the parameter values chosen) and vanishes in the core region of the 
diwhirl.  So the picture that emerges is this: the diwhirl pattern is 
robust, and
it coexists with, but is spatially distinct from, the oscillatory finite
wavelength pattern arising from the linear instability.

 By estimating the work done by a
 fluid element in a single  flow cycle (as seen in a Lagrangian
 reference frame), Groisman and Steinberg~\cite{steinberg98} argued that long
 wavelength stationary structures which exhibited significant
 asymmetry between inflow and outflow were possible in Couette
 flow. While this argument shows that diwhirl structures are
 physically plausible, it does not explain the mechanism by which they
 are sustained. Having the complete velocity and stress field
 available to us, we propose a more complete  mechanism.
  Figure 3 shows a vector plot
 of $\boldv$ near the outer cylinder at the center of the diwhirl
 structure. We see that the $v_\theta$
 field at the 
 diwhirl center is locally parabolic near the outer cylinder, 
 similar to the velocity field in the upper half of the channel in
 Dean flow, which displays a stationary vortex
 instability due to the corresponding unstable stratification of the
 azimuthal normal stress~\cite{shaqfeh91}. We therefore propose  the
 following mechanism for the 
 instability: a local, finite amplitude perturbation near the outer
 cylinder creates a locally parabolic velocity profile. This velocity
 profile results in an unstable stratification of hoop
 stress, just as in Dean flow, which drives inward radial motion. The
 fluid accelerates azimuthally as it moves radially inward, due to the
 base state velocity gradient, thus maintaining the increase in
 $v_\theta$, resulting in a self-sustaining process
 (figure 4). 

To summarize, we have computed stationary nontrivial solutions in
circular Couette flow using a FENE-P model by numerical continuation
from stationary bifurcations in Couette-Dean flow.  These solutions
show very strong localization, exist only for 
large values of $b$ and large
wavelengths, and show a hysteretic character in $\Wet$. The self-sustaining
mechanism  is related to the mechanism of instability in
viscoelastic Dean flow, and arises from a finite amplitude
perturbation giving rise to a locally parabolic profile of the
azimuthal velocity  near the upper
wall. The computed flow structures  are very similar to the experimentally
observed diwhirl patterns~\cite{steinberg97}. Along with the
solutions arising from the linear instability of the circular Couette flow base
state, we propose that these solutions form  building blocks for
spatiotemporal dynamics in the flow of elastic liquids. 

Financial support from NSF and ACS is gratefully acknowledged.
We would also like to thank Prof. Bamin Khomami,
Prof. R. Sureshkumar, and their research associates  for helpful discussions.

\bibliographystyle{unsrt}

\begin{thebibliography}{10}

\bibitem{denn76}
C.~J.~S. Petrie and M.~M. Denn.
\newblock {\em AIChE J.}, 22(2):209--236, 1976.

\bibitem{larson92}
R.~G. Larson.
\newblock {\em Rheol. Acta}, 31:213--263, 1992.

\bibitem{shaqfeh96rev}
E.~S.~G. Shaqfeh.
\newblock {\em Ann. Rev. Fluid Mech.}, 28:129--185, 1996.

\bibitem{lsm90}
R.~G. Larson, E~.S.~G. Shaqfeh, and S.~J. Muller.
\newblock {\em J. Fluid Mech.}, 218:573--600, 1990.

\bibitem{elata66}
H.~Rubin and C.~Elata.
\newblock {\em Phys. Fluids}, 9:1929--1933, 1966.

\bibitem{roisman69}
M.~M. Denn and J.~J. Roisman.
\newblock {\em AIChE J}, 15:454--459, 1969.

\bibitem{denn72}
Z.-S. Sun and M.~M. Denn.
\newblock {\em AIChE J}, 18:1010--1015, 1972.

\bibitem{giesekus72}
H.~Giesekus.
\newblock {\em Prog. Heat Mass Transfer}, 5:187--193, 1972.

\bibitem{joseph74}
G.~S. Beavers and D.~D. Joseph.
\newblock {\em Phys. Fluids}, 17:650--651, 1974.

\bibitem{datta64}
S.~K. Datta.
\newblock {\em Phys. Fluids}, 7:1915--1919, 1964.

\bibitem{walters64}
R.~H. Thomas and K.~Walters.
\newblock {\em J. Fluid Mech.}, 18:650--651, 1964.

\bibitem{lange00}
M.~Lange and B.~Eckhardt.
\newblock Preprint, 2000. Although these authors denote the solutions
  that they find as ``diwhirls'', their investigation is confined to
  regimes where $Re\gg 1$ and $\Wet/Re \ll 1$, where $Re$ is the Reynolds
  number. In these parameter regimes, fluid
  inertia plays a significant role. Thus, the patterns they simulate cannot
  be classified as being driven by elasticity alone.

\bibitem{steinberg97}
A.~Groisman and V.~Steinberg
\newblock {\em Phys. Rev. Lett.}, 78(8):1460--1463, 1997.

\bibitem{muller99}
B.~M. Baumert and S.~J. Muller.
\newblock {\em J. Non-Newtonian Fluid Mech.}, 83(1--2):33--69, 1999.

\bibitem{steinberg00}
A.~Groisman and V.~Steinberg.
\newblock {\em Nature}, 405:53--55, 2000.

\bibitem{dpl2}
R.~B. Bird, C.~F. Curtiss, R.~C. Armstrong, and O.~Hassager.
\newblock {\em Dynamics of polymeric liquids}, volume~2.
\newblock Wiley, New York, 2nd edition, 1987.

\bibitem{patera89}
Y.~Maday and A.~T. Patera.
\newblock In {\em State of the art surveys on computational mechanics}, pages
  71--143. ASME, 1989.

\bibitem{hughes82}
A.~N. Brooks and T.~J.~R. Hughes.
\newblock {\em Comp. Methods Appl. Mech. Eng.}, 32:199--259, 1982.

\bibitem{crochet87}
J.~M. Marchal and M.~J. Crochet.
\newblock {\em J. Non-Newtonian Fluid Mech.}, 26:77--114, 1987.

\bibitem{seydel94}
R.~Seydel.
\newblock {\em Practical bifurcation and stability analysis}.
\newblock Springer-Verlag, New York, 1994.

\bibitem{gmres}
Y.~Saad and M.~H. Schultz.
\newblock {\em SIAM J. Sci. Stat Comput.}, 7(3):856--869, 1986.

\bibitem{saad96}
Y.~Saad.
\newblock {\em Iterative methods for sparse linear systems}.
\newblock PWS Publishing Company, Boston, 1996.

\bibitem{graham00}
K.~A. Kumar and M.~D. Graham.
\newblock In preparation.

\bibitem{steinberg98}
A.~Groisman and V.~Steinberg.
\newblock {\em Phys. Fluids}, 10(10):2451--2463, 1998.

\bibitem{shaqfeh91}
Y.~L. Joo and E.~S.~G. Shaqfeh.
\newblock {\em Phys. Fluids A}, 3(7):1691--1694, 1991.

\end{thebibliography}
\newcommand{\noopsort}[1]{} \newcommand{\clarify}[1]{#1}

\newpage

\section*{Figure captions for Kumar and Graham}

\noindent {\bf Figure 1:} Density plot of $\alpha_{\theta\theta}$
(white is large  tension, black small) and contour plot
of the streamfunction at  $L=110.89$ ($\Wet=23.52$, $b=1830$,
 $S=1.2$, and $\epsilon=0.2$). For clarity, most of the flow domain is
not shown. Note the very strong localization of $\alpha_{\theta\theta}$
near the center. Away from the core, the structure is pure circular
Couette flow.

\vspace*{0.2in}

\noindent {\bf Figure 2:} Diwhirl solution amplitudes as functions of
$\Wet$ and $L$. Note that the curves at $L=9.11$ and $L=4.72$
 are very close together, while both curves are well separated from
 the curve at $L=3.08$ ($b=1830$, $S=1.2$, and
 $\epsilon=0.2$)

\vspace*{0.2in}

\noindent {\bf Figure 3:}Vector plot of  $\boldv$ near the outer
 cylinder at the 
 center of  the diwhirl structure (oblique arrows) and the base state (straight
arrows). The length of the arrows  is proportional to the magnitude
of the velocity. The axial velocity is identically zero in the base
state, and is zero by symmetry at the center of the diwhirl.

\vspace*{0.2in}
\noindent{\bf Figure 4:} Nonlinear self-sustaining mechanism
for the diwhirl patterns.

\end{document}